\documentclass[conference]{IEEEtran}
\IEEEoverridecommandlockouts

\usepackage[table,xcdraw]{xcolor}
\usepackage{pgf,tikz,pgfplots}
\usepackage{tikz-cd}

\usepackage{mdframed}
\mdfdefinestyle{MyFrame}{%
    linecolor=blue,
    outerlinewidth=4pt,
    roundcorner=20pt,
    innertopmargin=0.3mm,
    innerbottommargin=0.3mm,
    innerrightmargin=0.3mm,
    innerleftmargin=0.3mm,
    backgroundcolor=gray!20!white}

\pgfplotsset{compat=1.15}
\usetikzlibrary[patterns,arrows,shapes,shadows.blur,positioning,fit,backgrounds]

\usepackage{enumitem}
\setlist{nolistsep,leftmargin=*}

\usepackage{algorithm}
\usepackage{algpseudocode}

\usepackage{multirow}

\usepackage{soul}

\usepackage{fancybox,framed}

\usepackage{subcaption}
\usepackage{wrapfig}
\usepackage{tcolorbox}
\usepackage{booktabs} 
\usepackage{tabularx} 
\usepackage{lipsum}

\usepackage{mathtools,amssymb,amsthm}

\newtheorem{theorem}{Theorem}
\newtheorem{remark}{Remark}

\newtheorem{proposition}{Proposition}
\newtheorem{example}{Example}
\newtheorem{lemma}{Lemma}

\newtheorem{corollary}{Corollary}

\newcommand{\bC}{\mathbb{C}}

\newcommand{\bF}{\mathbb{F}}

\newcommand{\bN}{\mathbb{N}}

\newcommand{\bR}{\mathbb{R}}

\newcommand{\bZ}{\mathbb{Z}}

\newcommand{\cB}{\mathcal{B}}
\newcommand{\cC}{\mathcal{C}}

\newcommand{\cU}{\mathcal{U}}

\newcommand{\bolda}{\mathbf{a}}
\newcommand{\boldb}{\mathbf{b}}
\newcommand{\boldc}{\mathbf{c}}
\newcommand{\boldd}{\mathbf{d}}
\newcommand{\bolde}{\mathbf{e}}

\newcommand{\boldh}{\mathbf{h}}

\newcommand{\boldl}{\mathbf{l}}

\newcommand{\bolds}{\mathbf{s}}

\newcommand{\boldv}{\mathbf{v}}

\newcommand{\boldy}{\mathbf{y}}
\newcommand{\boldz}{\mathbf{z}}

\newcommand{\boldH}{\mathbf{H}}

\newcommand{\boldzero}{\boldsymbol{0}}

\DeclarePairedDelimiterX{\inp}[2]{\langle}{\rangle}{#1, #2}

\DeclareMathOperator{\Vor}{Vor}

\DeclarePairedDelimiter{\floor}{\lfloor}{\rfloor} 

\newcommand{\onenorm}[1]{\lVert #1 \rVert_1}
\newcommand{\twonorm}[1]{\lVert #1 \rVert_2}
\newcommand{\infnorm}[1]{\lVert #1 \rVert_\infty}

\DeclareSymbolFont{bbold}{U}{bbold}{m}{n}
\DeclareSymbolFontAlphabet{\mathbbold}{bbold}
\newcommand{\1}{\mathbbold{1}}

\DeclareSymbolFontAlphabet{\mathcal}{symbols}

\usepackage{csquotes}
\usepackage[%
backend=biber,
style=numeric,
citestyle=numeric,
sortcites=true,
giveninits=true,
maxcitenames=1,
maxbibnames=6,
minbibnames=6,
doi=false,
url=true,
maxnames=99,
hyperref=true,
isbn=false
]{biblatex}
\AtEveryBibitem{\clearfield{pages}}
\renewrobustcmd*{\bibinitdelim}{\,}
\renewrobustcmd*{\multicitedelim}{,\,}
\DeclareFieldFormat
[article,inbook,incollection,inproceedings,patent,thesis,unpublished]
{title}{#1\isdot}

\usepackage{times}

\newif\ifFULL
\FULLtrue

\usepackage[font={scriptsize}]{caption}

\def\BibTeX{{\rm B\kern-.05em{\sc i\kern-.025em b}\kern-.08em
    T\kern-.1667em\lower.7ex\hbox{E}\kern-.125emX}}
\begin{document}

\title{Efficient Vector Symbolic Architectures\\from Histogram Recovery}

\author{%
	\IEEEauthorblockN{\textbf{Zirui (Ken) Deng} and \textbf{Netanel Raviv}}
	\IEEEauthorblockA{
		Computer Science and Engineering, Washington University in St. Louis\\
		\texttt{d.ken,netanel.raviv@wustl.edu}} 
}

\maketitle

\begin{abstract}
	Vector symbolic architectures (VSAs) are a family of information representation techniques which enable \textit{composition}, i.e., creating complex information structures from atomic vectors via binding and superposition, and have recently found wide ranging applications in various neurosymbolic artificial intelligence (AI) systems and hardware systems.
	Recently, \citeauthor*{raviv2024linear} proposed the use of random linear codes in VSAs, suggesting that their subcode structure enables efficient unbinding, while preserving the quasi-orthogonality that is necessary for neural processing. 
	Yet, random linear codes are difficult to decode under noise, which severely limits the resulting VSA's ability to support \textit{recovery}, i.e., the retrieval of information objects and their attributes from a noisy compositional representation. 
	
	In this work we bridge this gap by utilizing coding theoretic tools.
	First, we argue that the concatenation of Reed-Solomon and Hadamard codes is suitable for VSA, due to the mutual quasi-orthogonality of the resulting codewords (a folklore result). 
	Second, we show that recovery of the resulting compositional representations can be done by solving a problem we call \textit{histogram recovery}.
	In histogram recovery, a collection of~$N$ histograms over a finite field is given as input, and one must find a collection of Reed-Solomon codewords of length~$N$ whose entry-wise symbol frequencies obey those histograms.
	We present an optimal solution to the histogram recovery problem by using algorithms related to list-decoding, and analyze the resulting noise resilience. 
	Our results give rise to a noise-resilient VSA with formal guarantees regarding efficient encoding, quasi-orthogonality, and recovery, without relying on any heuristics or training, and while operating at improved parameters relative to similar solutions such as the Hadamard code.
\end{abstract}


\section{Introduction}
The two leading approaches in AI are \textit{symbolic-AI}, in which concepts are represented as discrete symbols and manipulated using logical operations, and \textit{connectionist-AI}, which processes real-valued vectors through networks of neurons and learns by adjusting their interconnections. 
Both approaches currently struggle with \textit{compositionality}, i.e., the ability to manipulate representation of complex objects: symbolic-AI struggles to scale, while connectionist-AI lacks clear boundaries between concepts due to their real-valued representations. 
To address these limitations, vector-symbolic representations (VSAs, a.k.a hyperdimensional computing) combine both approaches by representing concepts as atomic vectors which are composed via algebraic operations and processed via connectionist methods. 
VSAs are one part of a broader effort to facilitate symbolic reasoning in neural computation, often referred to as \textit{neurosymbolic AI}. 
Neurosymbolic AI recently featured as the first among six possible futures for AI research in a white paper by the Computing Community Consortium of the Computing Research Association~\cite{cra}.

Information representation in VSAs begins by fixing (usually random) real-valued atomic vectors to represent attributes (e.g., vectors~$blue,green$ for colors, and vectors~$circle,triangle$ for shapes), and objects are formed by performing a \textit{binding} operation~$\otimes$ which associates different atomic vectors to form objects (e.g.~$green\otimes circle$ for a green circle).
The resulting bound vectors are then superimposed via addition (e.g.,  $green\otimes circle+blue\otimes triangle$).
This superposition is a fixed-width representation of an input (e.g., a visual scene) which contains objects with these attributes. Importantly, this representation must not confuse which attribute belongs to which object, a phenomenon termed ``superposition catastrophe,'' see  e.g.~\cite{bowers2014neural}.

VSAs initially emerged in computational neuroscience as effective models of cognition~\cite{smolensky1990tensor,kanerva2014computing,plate2003holographic}, but recently found multiple applications ranging from AI hardware~\cite{zou2022eventhd,hassan2024advancing}, to machine learning and AI systems~\cite{verges2025classification,zhao2025bridging,ganesan2021learning,hersche2023neuro,kymn2024compositional}.
The latter include transformer architectures~\cite{alam2023recasting}, whose inner attention module is a quintessential example for the usefulness of VSAs, as it essentially implements binding and superposition of keys and values according to the principles of~\cite{smolensky1990tensor}.
Multiple different architectures were proposed~\cite{kleyko2022survey,kleyko2023survey,schlegel2022comparison}, and a theoretical framework was developed in~\cite{thomas2021theoretical}.

Constructing a VSA requires a definition of atomic vectors, alongside mechanisms for binding and superposition thereof.
The vectors associated with any given VSA must be quasi-orthogonal, which facilitates neurally feasible retrieval~\cite{thomas2021theoretical} and memorization via associative memories~\cite{steinberg2022associative}.
Furthermore, certain application domains (e.g.,~\cite{kent2020resonator,frady2020resonator,kymn2024compositional}) require a \textit{recovery} algorithm, which is given a superposition of bound vectors and recovers all attributes of all objects while avoiding superposition catastrophe (e.g., $green\otimes circle+blue\otimes triangle\mapsto \{(green,circle),(blue,triangle)\}$); recovery can be viewed as performing \textit{superposition}-recovery to identify the individual objects, followed by \textit{binding}-recovery to identify their attributes.
Recovery is essential, e.g., when training a neural network to map a visual scene to its compositional representation; recovery is then required in order to identify the various objects in the scene and their attributes (see~\cite[Fig.~3]{frady2020resonator} and~\cite{kymn2024compositional}).
It should also be noted that recovery is the first open problem mentioned in the influential survey~\cite{kleyko2023survey} that should be addressed to built truly intelligent systems.

For the most common case of binding via point-wise product, a heuristic approach for recovery was taken in~\cite{frady2020resonator,kent2020resonator} using multiple Hopfield networks that operate in parallel, and yet the performance was rather limited, and no theoretical guarantees are known as of yet (similar solutions were explored in~\cite{yeung2024self,yeung2026coupled}). 
A radically different approach was taken in~\cite{raviv2024linear}, outperforming~\cite{frady2020resonator,kent2020resonator}, where atomic vectors are chosen as codewords in a random binary linear code~$\cC$, seen as~$\pm1$ vectors, and hence binding via point-wise product is isomorphic to exclusive-OR (or~$\bF_2$-addition). 
Specifically, different attributes are identified by codewords in trivially intersecting subcodes of~$\cC$ 
(e.g.,~$\cC_{\text{color}},\cC_{\text{shape}}\subseteq\cC$ with~$\cC_{\text{color}}\cap\cC_{\text{shape}}=\{\boldzero\}$).
Since the binding operation implicitly implements addition over the binary field~$\bF_2$, and since the subcodes intersect trivially, binding-recovery (i.e., $green\otimes circle\mapsto(green,circle)$) can be done via Gaussian elimination over~$\bF_2$. 
However, superposition-recovery in~\cite{raviv2024linear} is more involved, and polynomial run-time is not guaranteed in all cases.
Furthermore, in cases where the superposition is contaminated by noise, superposition-recovery of even a single codeword in a random binary linear code is known as NP-hard~\cite{berlekamp1978inherent}.

In this work we take the linear-codes based VSA of~\cite{raviv2024linear} further by presenting an explicit code construction with guaranteed quasi-orthogonality and an efficient recovery algorithm.
The codes we present result from concatenation of a Reed-Solomon outer code with a Hadamard inner code, a well-known technique~\cite{alon1992simple,forney1965concatenated} with guaranteed quasi-orthogonality.
Since this code is linear, its subcode structure can be utilized for efficient binding-recovery exactly as in~\cite{raviv2024linear}.
For superposition-recovery we begin by recovering the inner Hadamard code via simple lattice decoding; this process produces entry-wise histograms of the field elements in the superimposed Reed-Solomon codewords. 
Then, to extract the codewords in the superposition, one must find a set of Reed-Solomon codewords which obey these histograms, a problem we call \textit{histogram recovery}.
Aided by an algorithm for decoding Kautz-Singleton codes over a disjunctive channel due to~\cite{sidorenko2013low}, we present a solution to the histogram recovery problem, which cannot be generally improved. 
Crucially, our histogram recovery algorithm succeeds even if a bounded number of histograms are computed incorrectly. 
This superposition-recovery algorithm completes the picture of our VSA, which to the best of our knowledge is the first of its kind to formally guarantee efficient encoding, quasi-orthogonality, and efficient recovery.

\section{Preliminaries and Problem Definition}
The goal of vector-symbolic architectures is representing discrete symbols using real-valued vectors in a way which enables efficient neural processing of their compositions.
The problem setting includes a universe of all possible objects, where each object is uniquely identified by the values of its~$a$ attributes.
We identify the values of all attributes as discrete sets~$\cU_1,\ldots,\cU_a$ (e.g.,~$\cU_{\text{color}},\cU_{\text{shape}}$, etc), and the universe of all objects is their Cartesian product~$\cU=\cU_1\times\ldots\times\cU_a$. 

Endowing~$\cU$ with vector-symbolic representation requires identifying each attribute in~$\cup_{i=1}^a\cU_a$ as an \textit{atomic vector} in~$\bC^n$.
To represent objects, a binding operator~$\otimes$ is introduced.
An object with attributes~$\ell_1\in\cU_1,\ldots,\ell_a\in\cU_a$, which are represented by the atomic vectors~$\boldv_{1,\ell_1},\ldots,\boldv_{a,\ell_a}$, respectively, is represented by the bound vector~$\boldv_{1,\ell_1}\otimes\ldots\otimes\boldv_{a,\ell_a}$.
Then, bound vectors corresponding to different objects are superimposed via addition.
In this work we focus on the popular binding operation of point-wise product\footnote{
	Point-wise product binding is normally used in conjunction with~$\pm1$ atomic vectors. 
	Yet, recent findings~\cite{hiratani2023optimal} suggest that some optimal quadratic binding mechanisms of vectors in~$\bR^n$ implicitly implement point-wise products over~$\bC^{n/2}$. 
	Hence, for generality and simplicity of notation, we consider atomic vectors in~$\bC^n$ that are bound via point-wise product.}.
As pointed out in~\cite{thomas2021theoretical}, bound vectors must present some quasi-orthogonality in order to be neurally processed, i.e., the (normalized) inner product between any two distinct object representations must be close to zero.
A natural choice for atomic vectors is columns of a Hadamard matrix~\cite{liu2025linearithmic}, yet the resulting rate is very low.

The proposition of~\cite{raviv2024linear} is fixing a~$\mu$-incoherent~$[n,\log_p|\cU|]_p$ linear code~$\cC$ of length~$n$ and dimension~$\log_p|\cU|$ over some prime field~$\bF_p$ (represented using roots of unity of order~$p$).
That is, every distinct~$\boldc_1,\boldc_2\in\cC$ satisfy~$\frac{|\boldc_1\boldc_2^\dagger|}{\twonorm{\boldc_1}\twonorm{\boldc_2}}\le\mu $ for some small constant~$\mu$, where~$\cdot^\dagger$ denotes conjugate transposed.
Further, one fixes~$a$ trivially intersecting subcodes of~$\cC$ so that~$\cC=\cC_1\times \ldots\times \cC_a$, with~$\dim\cC_i=\log_p|\cU_i|$ for all~$i$.
The codewords of each~$\cC_i$ are used as atomic vectors for the attributes in~$\cU_i$.
The benefits are twofold: (a) the~$\mu$-incoherence of~$\cC$ is synonymous with the required quasi-orthogonality; and (b) the trivial intersection of the~$\cC_i$'s gives rise to binding-recovery (i.e., mapping $\boldv_{1,\ell_1}\otimes\ldots\otimes\boldv_{a,\ell_a}+\text{noise}\mapsto \{\boldv_{i,\ell_i}\}_{i=1}^a$) via mere decoding and Gaussian elimination over~$\bF_p$.

Our goal in this work is constructing a VSA which follows the premise of~\cite{raviv2024linear}, and also admits efficient \textit{superposition}-recovery algorithms. 
Since we focus on incoherent linear codes, their quasi-orthogonality and binding-recovery is an immediate corollary of~\cite{raviv2024linear}, and hence we focus only on constructing the entire code (Section~\ref{section:construction}) and its superposition-recovery~(Section~\ref{section:recovery}).
Figures of merit for evaluation are the code's rate and~$\mu$-incoherence, and the ability to recover a noisy superposition of as many codewords as possible in an efficient manner. 


\section{Concatenating Reed-Solomon and Hadamard}\label{section:construction}
As mentioned earlier, we introduce our VSA by following~\cite{raviv2024linear} and discussing the code~$\cC$ as a whole. 
In practice one should fix trivially intersecting subcodes~$\cC_1,\ldots,\cC_a$ for the atomic vectors of the various attributes according to the particular application in which VSA is deployed. 

Our construction follows a well-known technique of code concatenation between Reed-Solomon (over~$\bF_{p^m}$ for some~$m$ and a prime~$p$) and a Hadamard code (of size~$p^m$ over~$\bF_p$); we briefly present the full details for completeness. 
We begin by fixing an~$[N,K]_{p^m}$ Reed-Solomon code~$\cC_{\text{RS}}$ with~$N=p^m$, and fixing an isomorphism~$\phi$ between the additive group~$(\bF_p,+)$ of~$\bF_p$ and the group~$(\{\zeta^i\}_{i=0}^{p-1},\cdot)$, where~$\zeta\in\bC$ is a primitive root of unity of order~$p$.
Then, we fix a Hadamard code whose encoding function~$f_{\text{Had}}:\bF_p^m\to\bF_{p}^{p^m}$ maps a vector~$\bolda\in\bF_p^m$ to~$f_{\text{Had}}(\bolda)=(\bolda\boldb^\intercal)_{\boldb\in\bF_{p^m}}$, or equivalently, to the~$\bolda$'th column of a~$p^m\times p^m$ Hadamard matrix~$H$.
Our code~$\cC$ is obtained by considering all codewords in~$\cC_{\text{RS}}$, representing each~$\bF_{p^m}$ entry in each codeword as a vector in~$\bF_p^m$, mapping this~$\bF_p^m$ vector to a codeword in a Hadamard code via~$f_{\text{Had}}$, and then applying~$\phi$ on each of the resulting entries.
The resulting~$\cC$ is therefore a code of length~$p^{2m}$ and size~$p^{mK}$ over~$\bC$, given by
\begin{align}\label{equation:cDef}
	\cC=\{ (\phi(f_{\text{Had}}(c_1)),\ldots, \phi(f_{\text{Had}}(c_N)))\vert \boldc\in\cC_{\text{RS}} \},
\end{align}
where~$\phi$ acts in a point-wise manner.
\begin{proposition}\label{proposition:orthogonal}
	The code~$\cC$ is~$1-\frac{N-K+1}{N}$ incoherent. 
\end{proposition}
\ifFULL
\begin{proof}
	Let~$D=N-K+1$ be the minimum distance of~$\cC_{\text{RS}}$, and let~$\boldc_1,\boldc_2\in\cC$ be two distinct codewords in~$\cC$.
	First, observe that~$\twonorm{\boldc}=\sqrt{p^{2m}}$ for every~$\boldc\in\cC$, since all the entries of~$\boldc$ are roots of unity due to~$\phi$.
	Second, since the codewords of the Hadamard code are the columns of a~$p^m\times p^m$ Hadamard matrix~$H$, and since~$HH^\dagger=p^m I$, it follows that
	\begin{align*}
		\frac{1}{p^{2m}}|\boldc_1\boldc_2^{\dagger}|&=\frac{1}{p^{2m}}\left|\sum_{j\vert c_{1,j}= c_{2,j}}N\right|\le \frac{(N-D)N}{N^{2}}=1-\frac{D}{N}.\qedhere
	\end{align*}
\end{proof}
\fi
By Proposition~\ref{proposition:orthogonal}, the code $\cC$ is quasi-orthogonal and hence suitable for VSAs: for instance, by~\cite[Thm.~1]{raviv2024linear} it can support the neural retrieval of at least~$\frac{2N-D}{2N-2D}$ many items.
Furthermore, it is known that~$\cC$ is linear over~$\bF_p$~\cite[Ex.~10.2]{guruswami2018essential}, and hence it enables binding-recovery via~\cite{raviv2024linear}. 
In the next section we discuss superposition recovery of~$\cC$.

\section{Superposition Recovery Algorithm}\label{section:recovery}
Superposition recovery amounts to decoding~$\boldc_1,\ldots,\boldc_t$ from~$\bolds=\sum_{i=1}^t\boldc_t+\text{noise}$, where the~$\boldc_i$'s are not-necessarily-distinct codewords from~$\cC$. 
This is done in two steps.
First, each set of~$N=p^m$ consecutive entries of~$\bolds$ is processed separately. 
Each such set of entries contains the noisy superposition of Hadamard codewords, embedded in~$\bC^N$ (see~\eqref{equation:cDef}). 
These can be recovered efficiently since the codewords of the Hadamard code are an orthogonal basis to a lattice.
Recovering these Hadamard codewords results in an integer vector (a histogram) that indicates the frequencies of the Hadamard codewords that were superimposed in this particular set of entries of~$\bolds$.
This first step is done in Section~\ref{section:HadamardRecovery}.
Second, since the Hadamard codewords are in one-to-one correspondence with elements of the field~$\bF_{p^m}$, the resulting histograms in fact describe the frequencies of~$\bF_{p^m}$-symbols in the entries of the Reed-Solomon codewords that were used to construct the~$\boldc_i$'s (again, see~\eqref{equation:cDef}). 
The resulting \textit{histogram recovery} problem is solved in Section~\ref{section:HistoRecovery}.

\subsection{Recovery of Hadamard Codes via Lattice Decoding}\label{section:HadamardRecovery}
Let~$\bolds_i$ be the set of entries of~$\bolds$ indexed by~$(iN+1,\ldots,iN+N)$ for some~$i\in\{0,1,\ldots,N-1\}$.
According to~\eqref{equation:cDef}, $\bolds_i$ is a superposition of not-necessarily-distinct columns of a Hadamard matrix, plus a noise vector, i.e.,~$\bolds_i=H\boldh_i^\intercal+\bolde_i^\intercal$ for some~$\boldh_i\in\bZ_{\ge 0}^N$ and some~$\bolde_i\in\bC^n$. 


Therefore, extracting the histogram~$\boldh_i$ from~$\bolds_i$ amounts to decoding in the lattice~$H\bZ^{N}$ under complex valued noise; this requires the following characterization of \textit{Voronoi cells} in~$H\bZ^{N}$. 
For~$\boldh\in\bZ^{N}$ let~$\Vor(\boldh)=\{\boldy\in\bC^{N}|\boldh=\arg\min_{\boldz\in\bZ^{N}}\lVert H\boldz^\intercal-\boldy^\intercal \rVert_2\}$, i.e., all~$\bC^{N}$ points to which the closest lattice point is~$H\boldh^\intercal$ (in Euclidean distance). 

\begin{lemma}\label{lemma:VoronoiCell}
		For~$\boldh\in\bZ^{N}$ we have~$\Vor(\boldh)=H\boldh^\intercal+H(-0.5,0.5)^{N}+\iota H\bR^{N}$, where~$\iota=\sqrt{-1}$.
\end{lemma}
\ifFULL
\begin{proof}
	For any~$\boldy\in\bC^{N}$ and any~$\boldz\in\bZ^{N}$ we have that
	\begin{align*}
		\lVert \boldy^\intercal&-H\boldz^\intercal\rVert_2^2=(\boldy^\intercal-H\boldz^\intercal)^\dagger(\boldy^\intercal-H\boldz^\intercal)\\
		&=\lVert \boldy\rVert_2^2+N\lVert\boldz\rVert_2^2-2\boldz\Re( H^\dagger\boldy^\intercal)\\
		&=\lVert \boldy\rVert_2^2+ N\lVert\boldz-\textstyle\frac{1}{N}\Re(H^\dagger\boldy^\intercal)\rVert_2^2-\textstyle\frac{1}{N}\lVert\Re(H^\dagger\boldy^\intercal\rVert_2^2,
	\end{align*}
	where the last equality follows by completing the square.
	Therefore, the (unique) closest lattice point to a given~$\boldy\in\bC^{N}$ is~$H\boldh^\intercal$ whenever~$\boldh=\arg\min_{\boldz\in\bZ^{N}}\lVert\boldz-\frac{1}{N}\Re(H^\dagger\boldy^\intercal)\rVert_2^2$.
	Since~$\arg\min_{\boldz\in\bZ^{N}}\lVert\boldz-\frac{1}{N}\Re(H^\dagger\boldy^\intercal)\rVert_2^2$ is given by rounding $\frac{1}{N}\Re(H^\dagger\boldy^\intercal)$ to a nearest integer\footnote{If an entry is precisely between two integers, then the point does not belong to any Voronoi cell.}, for any~$\boldh\in\bZ^{N}$ we have
	\begin{align}\label{equation:Vorh}
		\Vor(\boldh)=\{\boldy\in\bC^{N}\vert \textstyle\frac{1}{N}\Re(H^\dagger\boldy^\intercal)\in\boldh^\intercal-(-0.5,0.5)^N \}.
	\end{align}
	It remains to show that the set in~\eqref{equation:Vorh} is equal to the one given in the statement of the lemma.
	In one direction, let~$\boldy\in\bC^N$ with~$\frac{1}{N}\Re(H^\dagger\boldy^\intercal)\in \boldh-(-0.5,0.5)^N$. 
	It follows that
	\begin{align*}
		\Re(H^\dagger\boldy^\intercal)&\in N\boldh^\intercal-(-N/2,N/2)^N\\
		H^\dagger\boldy^\intercal&\in N\boldh^\intercal-(-N/2,N/2)^N+\iota \bR^N\\
		\boldy^\intercal&\in H\boldh^\intercal+H(-0.5,0.5)^N+\iota H\bR^N,
	\end{align*}
	where the last transition follows by multiplying by~$\frac{1}{N}H$.
	In the other direction, for $\boldy^\intercal\in H\boldh^\intercal+H(-0.5,0.5)^N+\iota H\bR^N$ there exist~$\boldb\in(-0.5,0.5)^N$ and~$\boldd\in\bR^N$ such that~$\boldy^\intercal=H\boldh^\intercal+H\boldb^\intercal+\iota H\boldd^\intercal$, and hence
	\begin{align*}
		\textstyle\frac{1}{N}\Re(H^\dagger\boldy^\intercal)&=\textstyle\frac{1}{N}\Re(N\boldh^\intercal +N\boldb^\intercal+\iota N\boldd^\intercal)\\
		&=\boldh^\intercal+\boldb^\intercal\in \boldh^\intercal-(-0.5,0.5)^N.\qedhere
	\end{align*}
\end{proof}
\fi
Lemma~\ref{lemma:VoronoiCell} establishes both a unique decoding condition, and a simple decoding algorithm, as follows. 

\begin{corollary}\label{corollary:HadamardDecoding}
	If~$\bolde_i\in H(-0.5,0.5)^N+\iota H\bR^N$, then $\operatorname{round}(\frac{1}{N}\Re(H^\dagger\bolds_i^\intercal))=\boldh_i$, where~$\operatorname{round}$ denotes rounding.
\end{corollary}
\ifFULL
\begin{proof}
	It follows from Lemma~\ref{lemma:VoronoiCell} that~$\bolds_i\in\Vor(\boldh_i)$. 
	Furthermore, it follows from the Voronoi cell characterization in~\eqref{equation:Vorh} that rounding acts as a decoding algorithm.
\end{proof}
\fi
Notably, it follows from Corollary~\ref{corollary:HadamardDecoding} that the decoding algorithm $\operatorname{round}(\frac{1}{N}\Re(H^\dagger\bolds_i^\intercal))$ succeeds even for some \textit{unbounded} error vectors~$\bolde_i$. 
However, a norm bound on~$\bolde_i$ which guarantees successful decoding can be established as follows.
\begin{corollary}\label{corollary:L2boundHadamard}
	If~$\twonorm{\bolde_i}< \frac{\sqrt{N}}{2}$ then	$\operatorname{round}(\frac{1}{N}\Re(H^\dagger\bolds_i^\intercal))=\boldh_i$.
\end{corollary}
\ifFULL
\begin{proof}
	Write~$H^\dagger \bolde_i=\bolda+\iota \boldb$ with $\bolda,\boldb\in\bR^N$, hence~$\bolde_i=H\bolda^\intercal+\iota H\boldb^\intercal$, and observe that~$\bolds_i\in \Vor(\boldh_i)$ if and only if~$\bolda\in(-0.5,0.5)^N$ by Lemma~\ref{lemma:VoronoiCell}.
	Additionally,~$\twonorm{\bolde_i}^2=\twonorm{H\bolda^\intercal+\iota H\boldb^\intercal}^2=N(\twonorm{\bolda}^2+\twonorm{\boldb}^2)$.
	Therefore, since~$\twonorm{\bolde_i}<\sqrt{N}/2$, it follows that~$N(\twonorm{\bolda}^2+\twonorm{\boldb}^2)< N/4$, and in particular, that~$\twonorm{\bolda}< 1/2$.
	Since~$\infnorm{\bolda}< \twonorm{\bolda}$, it follows that~$\bolda\in(-0.5,0.5)^N$, hence~$\bolds_i\in\Vor(\boldh_i)$, and the claim follows by Corollary~\ref{corollary:HadamardDecoding}.
\end{proof}
\fi

\subsection{Histogram Recovery of Reed Solomon Codes}\label{section:HistoRecovery}
By inverting the function~$f_{\text{Had}}$, each histogram~$\boldh_i$ produced in Section~\ref{section:HadamardRecovery} can be seen as a histogram of~$\bF_{p^m}$ elements, and let~$\{\boldh_1,\ldots,\boldh_N\}$ be those histograms.
We now ought to find codewords in~$\cC_{\text{RS}}$ whose entry-wise frequencies obey~$\{\boldh_1,\ldots,\boldh_N\}$.
In the sequel our algorithm will induce a parameter tradeoff which guarantees that such codewords exist and are unique.
Furthermore, an example will be provided of a Reed-Solomon code whose parameters deviate from this tradeoff by~$1$, and contains two distinct collections of codewords which induce identical histograms, making histogram recovery information-theoretically impossible in this case (see Example~\ref{example:HRlimit}).
This example implies that our algorithm cannot be improved in general.
We also note that our algorithm is conducted over~$\bF_{p^m}$ in its entirety, and as such it is not prone to any numerical accuracy issues.

We rely on a decoding algorithm of Kautz-Singleton codes~\cite{kautz1964nonrandom} over a disjunctive channel due to~\cite{sidorenko2013low}. 
Kautz-Singleton codes are binary codes (i.e., codewords over~$\{0,1\}$) which result by taking a Reed-Solomon code of length~$N$ over some large finite field~$\bF_{p^m}$ and representing each~$\alpha\in\bF_{p^m}$ in every entry of every codeword  in its ``one-hot'' representation, i.e., a vector of length~$p^m$ which contains a~$1$ in the entry corresponding to~$\alpha$ and~$0$ elsewhere.
\ifFULL
That is, for any Reed-Solomon codeword~$\boldc=(c_{1},\ldots,c_{N})$ in a Reed-Solomon code of length~$N$ over~$\bF_{p^m}$, the corresponding Kautz-Singleton codeword~$W=(w_{j,\ell})_{j\in[p^m],\ell\in[N]}$ is defined by 
\begin{align}\label{eq:ksdef}
	w_{j,\ell} =
	\begin{cases}
		1, & c_{\ell}=\alpha_j,\\
		0, & c_{\ell}\neq \alpha_j,
	\end{cases},
\end{align}
where~$\alpha_1,\ldots,\alpha_{p^m}$ is any fixed ordering of the elements of~$\bF_{p^m}$.
\begin{example}\label{example:RStoKS}
	For~$p=7$ and~$m=1$ denote~$\bF_{p^m}=\bF_7=\{\alpha_1,\ldots,\alpha_7\}$ and let~$\boldc = (\alpha_2, \alpha_4, \alpha_6, \alpha_1, \alpha_2, \alpha_5)$ be some Reed-Solomon codeword. 
	The corresponding Kautz-Singleton codeword is:
	\begin{align*}
		W = \begin{bmatrix}
			0 & 0 & 0 & 1 & 0 & 0 \\
			1 & 0 & 0 & 0 & 1 & 0 \\
			0 & 0 & 0 & 0 & 0 & 0 \\
			0 & 1 & 0 & 0 & 0 & 0 \\
			0 & 0 & 0 & 0 & 0 & 1 \\
			0 & 0 & 1 & 0 & 0 & 0 \\
			0 & 0 & 0 & 0 & 0 & 0
		\end{bmatrix}.
	\end{align*}
\end{example}
\fi
Ref.~\cite{sidorenko2013low} studied the problem of decoding these code in a \textit{synchronous multiple access disjunctive channel}, where multiple users which possess different codewords transmit simultaneously to one receiver that receives the entry-wise Boolean OR of the codewords. 
We shall see next that histogram recovery can be performed by carefully using the algorithm of~\cite{sidorenko2013low} as a black box.
We begin by demonstrating histogram recovery for the special case of noiseless histograms formed from distinct codewords, and later progress to the general and noisy cases. 

\subsubsection{Noiseless histograms of distinct codewords}\label{section:noiselessKS}
Let~$\bar{\boldc}_1,\ldots,\bar{\boldc}_t\in\cC_{\text{RS}}$ be the Reed-Solomon codewords used to produce~$\boldc_1,\ldots,\boldc_t$ via~\eqref{equation:cDef}, and let~$W^{(1)},\ldots,W^{(t)}\in\{0,1\}^{p^m\times N}$ be the corresponding Kautz-Singleton codewords. 

Given the~$N$ \emph{noiseless} histograms~$\{\boldh_1,\ldots,\boldh_N\}$ arrange them into a matrix~$\boldH = [\boldh_1 \cdots \boldh_N] =(h_{i,j})_{i\in[p^m],j\in[N]}\in \bZ_{\ge 0}^{p^m\times N}$,
and let~$Y=(y_{i,j})_{i\in[p^m],j\in[N]} \in \{0,1\}^{p^m\times N}$ be such that~$y_{i,j} = 1$ if~$h_{i,j} > 0$ and~$y_{i,j} = 0$ otherwise. 
It is readily verified that~$Y$ is the element-wise Boolean OR of~$W^{(1)}, \ldots, W^{(t)}$, 
i.e., $Y \triangleq W^{(1)} \vee \cdots \vee W^{(t)}$. 
We utilize a Kautz-Singleton decoding algorithm~\cite[Alg.~2]{sidorenko2013low}, which is inspired by the well-known Guruswami-Sudan list decoding algorithm~\cite[Sec.~2.2]{guruswami1998improved}. 
It uses~$M \triangleq dY$ as input, where~$d$ is some tunable positive integer, and outputs a list of \emph{unique} Kautz-Singleton (equivalently Reed-Solomon) codewords, whose size depends on~$d$. 

According to~\cite[Thm.~1]{sidorenko2013low}, a Kautz-Singleton codeword~$W$ is in the list of outputs given input~$M$ if
\begin{align}\label{eq:innerproductcondition}
	\langle W, M \rangle \ge \sqrt{2(K-1)c(M)},
\end{align} 
where the entries of~$W$ and~$M$ are considered as integers in the inner product~$\langle W, M \rangle \triangleq \sum_{i=1}^{p^m} \sum_{j=1}^N w_{i,j}m_{i,j}$, and~$c(M) \triangleq \frac{1}{2}\sum_{i=1}^{p^m} \sum_{j=1}^N m_{i,j} (m_{i,j}+1)$ is called the \emph{cost} of~$M$. 
Now, if~$W \in \{W^{(i)}\}_{i=1}^t$, then~$\langle W, M \rangle = dN$.
On the other hand, since the sum of every column of~$\boldH$ equals~$t$, it follows that~$c(M) \le \frac{1}{2}d(d+1)t N$.
Therefore, it follows from~\eqref{eq:innerproductcondition} that all codewords~$\{W^{(i)}\}_{i=1}^t$ are included in the output of~\cite[Alg.~2]{sidorenko2013low}~if
\begin{align}\label{eq:noiselessellupperbound1}
	dN &\ge \sqrt{d(d+1)(K-1)t N} \mbox{, i.e.,}\nonumber\\
	t &\le \frac{d}{d+1}\frac{N}{K-1}. 
\end{align} 
Taking~$t$ which satisfies~\eqref{eq:noiselessellupperbound1}, therefore, guarantees that all ``correct'' codewords~$W^{(1)},\ldots, W^{(t)}$ are contained in the list\footnote{For~$t = \frac{d}{d+1}\frac{N}{K-1}$, the output list size is upper bounded by~$(d+1)t$~\cite[Cor.~1]{sidorenko2013low}, and yet it is pointed out in~\cite{sidorenko2013low} that this bound is often loose in practice.}. 
However, the list will also contain ``spurious'' codewords which are neither one of $W^{(1)},\ldots, W^{(t)}$.
In what follows it is shown that these spurious codewords can be filtered out according to the value of their inner product with~$M$.
To this end, we provide the following lemma.


\begin{lemma}\label{lemma:ellkminusone}
	For any Kautz-Singleton codeword $\bar{W}\notin\{W^{(i)}\}_{i=1}^t$, we have $|\operatorname{supp}(\bar{W}) \cap \operatorname{supp}(Y)| \le t(K-1)$, where~$\mathrm{supp}(\cdot)$ is the set of nonzero entries (row-column pairs).
\end{lemma}
\ifFULL
\begin{proof}
	Since any two distinct Reed-Solomon codewords agree in at most~$K-1$ entries, it follows that 
	$|\mathrm{supp}(\bar{W}) \cap \mathrm{supp}(W^{(i)})| \le K-1$ for each~$i\in[t]$.
	Therefore, 
	\begin{align*}
		\bigl\lvert \mathrm{supp}(\bar{W}) &\cap \mathrm{supp}(Y)\bigr\rvert \;=
		\bigl\lvert \mathrm{supp}(\bar{W}) \cap \bigcup_{i=1}^\ell \mathrm{supp}(W^{(i)})\bigr\rvert \\
		&\le
		\sum_{i=1}^\ell
		\bigl\lvert \mathrm{supp}(\bar{W}) \cap \mathrm{supp}(W^{(i)})\bigr\rvert\le t(K-1). \qedhere
	\end{align*}
\end{proof}
\fi
It follows from Lemma~\ref{lemma:ellkminusone} that every Kautz-Singleton codeword~$\bar{W}\notin\{W^{(i)}\}_{i=1}^t$ satisfies~$\langle \bar{W}, M \rangle \le dt(K-1)$. 
Hence, by requiring that $dt(K-1) < dN$, we effectively guarantee that all spurious codewords yield an inner product with~$M$ of strictly less than~$dN$, and hence can be filtered out. 
Since the latter condition is equivalent to~$t<\frac{N}{K-1}$, along with~\eqref{eq:noiselessellupperbound1}, it follows that
\begin{align}\label{eq:noiselessellupperbound}
	t\le \operatorname{min}\left\{\left\lfloor\frac{d}{d+1}\frac{N}{K-1}\right\rfloor, \left\lfloor\frac{N-1}{K-1}\right\rfloor\right\},
\end{align}
suffices to recover the original~$t$ distinct Reed-Solomon codewords~$\{\bar{\boldc}_i\}_{i=1}^{t}$. 
That is, apply the Kautz-Singleton decoding algorithm~\cite[Alg.~2]{sidorenko2013low} on~$M$ to obtain a polynomial-size list of candidate codewords. 
Then, reject any codeword~$\bar{W}$ such that~$\langle\bar{W},M\rangle<dN$, and output the remaining codewords. 
\ifFULL
\begin{remark}	
	We clarify the role of the parameter~$d$ in the above analysis.
	On the one hand, increasing~$d$ to~$d=N-1$ provides the maximum value of~$t$ in~\eqref{eq:noiselessellupperbound}, i.e., the maximum number of distinct codewords that can be recovered given their histograms. 
	On the other hand, increasing~$d$ leads to a larger output list (more precisely, a weaker upper bound), which requires more time in order to filter out the correct superimposed codewords.
	Specifically, the size of the output list with~$M$ as input has an upper bound of~$\sqrt{\frac{2c(M)}{K-1}}$~\cite{koetter2003algebraic},
	which is upper bounded by~$\sqrt{\frac{d(d+1)t N}{K-1}} \le \frac{dN}{K-1}$ since~$c(M)\le\frac{1}{2}d(d+1)tN$ and due to~\eqref{eq:noiselessellupperbound}.
	Nevertheless, the list size remains polynomial (specifically quadratic) in~$N$ for the case~$d=N-1$.
\end{remark}
\fi

Finally, we demonstrate the optimality of the above decoding. 
The following example shows that it is in general not possible to perform histogram recovery for~$t>\floor{\frac{N-1}{K-1}}$ codewords (this is called the \emph{strength} of a Kautz-Singleton, see~\cite{kautz1964nonrandom,sidorenko2013low}).
\begin{example}\label{example:HRlimit}
	Consider a Reed-Solomon code over~$\bF_7$ (i.e.,~$p=7$ and~$m=1$) of length~$N=7$ and dimension~$K=2$, spanned by~$\1$ and~$(0,1,\ldots,6)$, whose strength is~$\floor{\frac{7-1}{2-1}}=6$.
	The following two collections of~$7$ codewords each are both contained in the code, and induce identical histograms:
	\begin{align*}
		\{\alpha\cdot \1\}_{\alpha\in\bF_7}\mbox{ and }	\{ \alpha\cdot\1+(0,1,\ldots,6) \}_{\alpha\in\bF_7}.
	\end{align*}
	Therefore, it is impossible to distinguish between the cases that the original superimposed codewords are one or the other.
\end{example}

\subsubsection{Noisy histograms of distinct codewords}\label{section:noisyhist}
Now suppose some histograms are computed incorrectly. 
Let~$\boldh_1',\ldots, \boldh_N'$ be the noisy histograms, and arrange them into a matrix~$\boldH'=(h'_{i,j})_{i\in[p^m],j\in[N]}=[\boldh_1'\cdots \boldh_N']$.
To define the bounded error patterns in this section, let
\begin{align}\label{equation:e1e2}
	e_1(\boldH,\boldH')&\triangleq |\{j\in[N]:\boldh_j\ne\boldh'_j\}|,\nonumber\\
	e_2(\boldH,\boldH')&\triangleq \max\{d_1(\boldh_j,\boldh'_j)\}_{j\in[N]}.
\end{align}
\ifFULL
In words, $e_1(\boldH, \boldH')$ denotes the number of incorrect histograms, while~$e_2(\boldH, \boldH')$ denotes the maximum deviation of any incorrect histogram from its correct counterpart, measured by~$\ell_1$-distance. 
\fi
Below we provide an algorithm which outputs the correct~$\bar{\boldc}_1,\ldots,\bar{\boldc}_t$, assuming they are distinct (the general case is treated in Section~\ref{section:notdistinct}), and assuming that
$e_1(\boldH,\boldH')\le s$ and~$e_2(\boldH,\boldH')\le r$ for some nonnegative integers~$r$ and~$s$ such that~$(r+1)s < N$.
The maximum value of~$t$ which enables this noise-resilient histogram recovery is bounded by some function of~$r$ and~$s$ that is described shortly. 

We first observe that~$t$ need not be known a priori---since\footnote{
	Notice that if the~$e_1$ quantity is positive then there are two columns whose~$d_1$ distance is positive, and hence the~$e_2$ quantity is positive as well.
	Consequently, the limitation~$(r+1)s<N$ implies that~$s< N/2$ since otherwise if~$s\ge N/2$ then~$r\ge 1$, and hence~$(r+1)s\ge N$.
}~$s<N/2$, the correct value of~$t$ can be deduced by taking the majority value over the entry sums of the columns of~$\boldH'$.
With this in mind, we pre-process~$\boldH'$ by first taking a majority vote among the column sums of~$\boldH'$ to obtain~$t$, and then set to zero the columns of~$\boldH'$ which do not sum to~$t$.
Let~$e$ denote the number of zeroed columns, and observe that~$0\le e\le s$.

Similar to Section~\ref{section:noiselessKS}, let~$Y' \in \{0,1\}^{p^m\times N}$ be such that~$y'_{i,j} = 1$ if~$h'_{i,j} > 0$ and~$y'_{i,j} = 0$ otherwise, and let~$M' \triangleq dY'$.
Observe that since~$e_1(\boldH,\boldH')\le s$, it follows that~$\langle W^{(i)}, M' \rangle \ge d(N-s)$ for every originally superimposed codeword~$W^{(i)}$. 
Also, since~$e$ columns have been zeroed in pre-processing, it follows that~$c(M') \le \frac{1}{2}d(d+1)t(N-e)$.
Therefore, it follows from~\eqref{eq:innerproductcondition} that if
\begin{align}\label{eq:noisyellupperbound1}
	d(N - s) &\ge \sqrt{d(d+1)(K-1)t(N-e)}\mbox{, i.e.,}\nonumber\\
	t &\le \frac{d}{d+1} \cdot\frac{(N-s)^2}{(K-1)(N-e)},
\end{align}
then the correct codewords are in the output list.
To identify these~$W^{(i)}$'s from the output, we present the following lemma.

\begin{lemma}\label{lemma:hammingandellone}
	Let \(\boldH, \boldH' \in \bN^{p^m\times N}\) and~$Y,Y'\in\{0,1\}^{p^m\times N}$ as described above.
	Then, for each column index~$j\in[N]$ we have that~$d_H(Y_j,Y'_j) \le d_1(\boldh_j,\boldh'_j)$, where~$Y_j$ (resp.~$Y'_j$, $\boldh_j$, $\boldh_j'$) denotes the~$j$'th column of~$Y$ (resp.~$Y'$, $\boldH$, $\boldH'$). 
\end{lemma}
\ifFULL
\begin{proof}
	Fix $j\in[N]$ and observe that $d_H(Y_j,Y'_j) =\sum_{i=1}^{p^m} \1(y_{i,j}\neq y'_{i,j})$, where~$\1(\cdot)$ is the zero-one indicator of the given event.
	Since~$y_{i,j}\neq y'_{i,j}$ precisely when one of~$h_{i,j},h'_{i,j}$ is zero and the other is positive, in which case~$\lvert h_{i,j}-h'_{i,j}\rvert \ge 1$, it follows that $\1(y_{i,j}\neq y'_{i,j}) \le |h_{i,j}-h'_{i,j}|$.
	Summing over~$i\in \{1,\dots,p^m\}$ yields
	\begin{align*}
		d_H(Y_j,Y'_j)&=\textstyle\sum_{i=1}^{p^m}\1(y_{i,j}\neq y'_{i,j}) \\
		&\le \textstyle\sum_{i=1}^{p^m} |h_{i,j}-h'_{i,j}|
		=d_1(\boldh_j,\boldh'_j). \qedhere
	\end{align*}
\end{proof}
\fi
Lemma~\ref{lemma:hammingandellone} readily implies the following bound on the inner product between any codeword~$\bar{W}\notin\{W^{(i)}\}_{i=1}^t$ and the erroneous~$M'$.
\begin{lemma}
	Every Kautz-Singleton codeword $\bar{W}\notin\{W^{(i)}\}_{i=1}^t$ satisfies $\langle \bar{W}, M' \rangle \le d(t(K-1)+r(s-e))$.
\end{lemma}
\ifFULL
\begin{proof}
	Let~$C\subseteq[N]$ be the set of indices~$j\in[N]$ such that~$\boldh_j=\boldh_j'$, let~$Z_1\subseteq[N]$ be the set of~$e$ indices of erroneous columns of~$\boldH'$ that were zeroed during pre-computation, and let~$Z_2$ be the set of at most~$s-e$ indices of erroneous columns of~$\boldH'$ that were not zeroed during pre-computation.
	We have that
	\begin{align*}
		&\langle \bar{W}, M' \rangle \;=\sum_{j=1}^{N}\bar{W}_j(M_j')^\intercal\\
		&=\sum_{j\in C}\bar{W}_j(M_j')^\intercal +\sum_{j\in Z_1}\bar{W}_j(M_j')^\intercal+\sum_{j\in Z_2}\bar{W}_j(M_j')^\intercal\\
		&\overset{(a)}{=} \sum_{j\in C}\bar{W}_jM_j^\intercal +\sum_{j\in Z_2}\bar{W}_j(M_j')^\intercal\\
		&\overset{(b)}{\le} \sum_{j\in C}\bar{W}_jM_j^\intercal +\sum_{j\in Z_2}(\bar{W}_j(M_j)^\intercal+dr)
		\\
		&= \sum_{j\in C\cup Z_2}\bar{W}_j(M_j)^\intercal +(s-e)dr\\
		&\overset{(c)}{\le} \sum_{j\in [N]}\bar{W}_j(M_j)^\intercal +(s-e)dr\\
		&\overset{(d)}{=}  d(t(K-1) +(s-e)r)
	\end{align*}
	where~$(a)$ follows by the definition of~$C$ and~$Z_1$;
	$(b)$ follows from Lemma~\ref{lemma:hammingandellone} since~$d_1(\boldh_j,\boldh_j') \le r$ implies that~$d_H(M_j,M'_j) = d_H(Y_j,Y'_j) \le r$, and since all entries of~$M$ and~$M'$ are either~$0$ or~$d$; $(c)$ follows since all entries of~$\bar{W}$ and~$M$ are nonnegative; and~$(d)$ follows from Lemma~\ref{lemma:ellkminusone}.
\end{proof}
\fi

Therefore, every correct codeword's inner product with~$M'$ is at least~$d(N-s)$, whereas every incorrect codeword's inner product with~$M'$ is at most~$d(t(K-1)+r(s-e))$.
Hence, requiring that the former is strictly larger than the latter leads to
\begin{align}\label{eq:noisyellupperbound2}
	t < \frac{N-s-r(s-e)}{K-1}.
\end{align}
Combining~\eqref{eq:noisyellupperbound1} and~\eqref{eq:noisyellupperbound2} gives us the following bound: 
\begin{align}\label{eq:bothnoisesbound}
	t\le \min\bigg\{\left\lfloor\frac{d}{d+1}\frac{(N-s)^2}{(N-e)(K-1)}\right\rfloor,\nonumber\\ \left\lfloor\frac{N-s-r(s-e)-1}{K-1}\right\rfloor\bigg\}
\end{align}

Therefore, even in the case of noisy histograms, successful decoding is still possible as long as~\eqref{eq:bothnoisesbound} holds: apply the Kautz-Singleton decoding algorithm~\cite[Alg.~2]{sidorenko2013low} on~$M'$ to obtain a polynomial-size list of candidate codewords. 
Then, reject any codeword~$\bar{W}$ such that~$\langle \bar{W},M'\rangle<d(N-s)$, and output the remaining codewords.
Notice that an estimate of~$s$ is required (in the worst case assume~$s=\floor{(N-1)/2}$), 
whereas the parameters~$r$ and~$e$ need not be known. 
In addition, if~$s=e=0$, then the bound in~\eqref{eq:bothnoisesbound} specifies to the one in~\eqref{eq:noiselessellupperbound}, indicating that the noiseless problem in Section~\ref{section:noiselessKS} is a special case of the noisy one discussed herein. 

\ifFULL
The next remark shows the benefit of pre-processing, as well as the optimality of the above Kautz-Singleton decoding in the presence of erasure noise.
\begin{remark}\label{remark:benefitoferasure}
	The bound in~\eqref{eq:bothnoisesbound} implies that identifying columns to be erased in pre-processing is potentially beneficial in the sense that it leads to a higher upper bound on~$t$ when~$e > 0$.
	In particular, when~$e = s$, i.e., when all noisy columns are erased during pre-processing, the corresponding bound is
	\begin{align*}
		t\le \min\left\{\left\lfloor\frac{d}{d+1}\frac{N-e}{K-1}\right\rfloor, \left\lfloor\frac{N-e-1}{K-1}\right\rfloor\right\},
	\end{align*}
	which reduces to~$t\le \left\lfloor\frac{N-e-1}{K-1}\right\rfloor$ when~$d = N - e - 1$.
	This upper bound on~$t$ is the strength of the Kautz-Singleton code of length~$N-e$ and dimension~$K$. 
	Therefore, the above Kautz-Singleton-based decoding is optimal when all errors in the histograms are erasures.
	In other words, there exists no decoding algorithm that outperforms our decoding for \textit{every} parameter regime.
\end{remark}
\fi

\subsubsection{Noisy histograms of not-necessarily-distinct codewords}\label{section:notdistinct}
\ifFULL
We now consider the case where the Reed-Solomon codewords~$\bar{\boldc}_1,\ldots,\bar{\boldc}_t$ are not necessarily distinct.
Let~$t$ denote the \textit{total} number of codewords (including multiplicities), and let~$0 < u \le t$ be the number of \textit{unique} codewords;
we proceed to focus on the general case with noisy histograms.

From Section~\ref{section:noiselessKS} and Section~\ref{section:noisyhist}, duplicate codewords do not change any entries in~$Y'$ (or~$Y$ in the noiseless case), and given~$M' = dY'$ as input, Kautz-Singleton decoding will return only the distinct codewords.
To find their multiplicities, subtract the histograms of these unique codewords given as output from the histograms given as input, and run the decoding algorithm again.
Keep repeating this process until the residual histograms become zero, at which point the iterative process terminates, and the multiplicities will be available from the histograms given throughout all iterations.


It remains to be argued that the assumed error bounds on~$e_1$ and~$e_2$ in~\eqref{equation:e1e2} are maintained at every step.
Indeed, let~$\boldh_i^{(j)}$ be the noiseless histogram of the~$i$-th coordinate at step~$j$, with~$\boldh_i^{(0)} \triangleq \boldh_i$, let~$\boldh_i^{(j)\prime}$ be the potentially noisy histogram of the~$i$-th coordinate at step~$j$, with~$\boldh_i^{(0)\prime} \triangleq \boldh_i^\prime$, and let~$\boldl_i$ be the histogram of the~$u$ distinct codewords at step~$i$.
Notice that if~$\boldh'_i = \boldh_i$ for some~$i \in [N]$, i.e., if the~$i$-th histogram is noiseless, then we have~$$\boldh_i^{(1)} = \boldh_i^{(0)} - \boldl_i = \boldh_i^{(0)\prime} - \boldl_i = \boldh_i^{(1)\prime},$$ and subsequently~$$\boldh_i^{(j+1)} = \boldh_i^{(j)} - \boldl_i = \boldh_i^{(j)\prime} - \boldl_i = \boldh_i^{(j+1)\prime}$$ at every step. 
In other words, noiseless histograms remain noiseless throughout. On the other hand, if~$0<d_1(\boldh_i, \boldh'_i) \le r$ for some~$i \in [n]$, we still have
\begin{align*}
	d_1(\boldh_i^{(j+1)\prime}, \boldh_i^{(j+1)}) &= d_1(\boldh_i^{(j)\prime} - \boldl_i, \boldh_i^{(j)} - \boldl_i)\\ &= d_1(\boldh_i^{(j)\prime}, \boldh_i^{(j)}) \le r.	
\end{align*}

It is possible that at some step~$j$ we observe some~$\boldh_i^{(j)\prime}$ with at least one negative entry. If this is the case, then this histogram is noisy, and we zero the corresponding entries in~$\boldH'$ as in the pre-computation step in Section~\ref{section:noisyhist}.
Let~$0 \le e_g \le s$ be the total number of zeroed columns throughout all steps; notice that in the case of distinct codewords we have~$e_g = e$. 
From~\eqref{eq:bothnoisesbound} we then have that as long as
\begin{align}\label{eq:boundonu}
	u \le \min\left\{\left\lfloor\frac{d}{d+1}\frac{(N-s)^2}{(N-e_g)(K-1)}\right\rfloor, \left\lfloor\frac{N-s-r(s-e_g)-1}{K-1}\right\rfloor\right\},
\end{align}
successful histogram recovery is possible.
\begin{remark}
	There is \emph{no restriction} on the number~$t$ of superimposed codewords, as long as the number~$u$ of \emph{distinct} codewords obeys~\eqref{eq:boundonu}.
\end{remark}
\fi
\ifFULL
\subsubsection{Summary}
We have thereby proved the following theorem.
\fi
\begin{theorem}\label{thm:distinctbound}
	Let~$\boldH = [\boldh_1 \cdots \boldh_N]$ be noiseless histograms of some~$t$ not necessarily distinct codewords of a Reed-Solomon code of length~$N$ and dimension~$K$, of which~$u \le t$ are distinct, and let~$\boldH' = [\boldh'_1 \cdots \boldh'_N]$ be possibly noisy histograms such that~$e_1(\boldH,\boldH')\le s$ and~$e_2(\boldH,\boldH')\le r$ for some nonnegative integers~$r$ and~$s$ with~$(r+1)s < N$. 
	Then, for any positive integer~$d$,  our algorithm with~$d$ and~$\boldH'$ as input returns the correct~$t$ codewords (i.e., correct codewords with correct multiplicities) as long as~$u \le \min\left\{\left\lfloor\frac{d(N-s)^2}{N(d+1)(K-1)}\right\rfloor, \left\lfloor\frac{N-s-rs-1}{K-1}\right\rfloor\right\}$.
\end{theorem}
\ifFULL
Namely, as long as the number~$u$ of distinct Reed-Solomon codewords is~$O(\frac{N}{K})$, and the error in computing the~$N$ histograms from the codewords is bounded~\eqref{equation:e1e2}, recovering the codewords and their multiplicities can be done efficiently.
Moreover, given this restriction on the number of distinct codewords, the \textit{total} number of codewords~$t$ in superposition is not bounded.
Since in each iteration of the algorithm in Section~\ref{section:notdistinct} at least one codeword from the superposition is found, in the worst case~$t$ iterations of the algorithm in Section~\ref{section:notdistinct} will provide all codewords in the superposition.
\fi

\begin{figure}[htbp]
	\centering
	\includegraphics[width=0.4\textwidth]{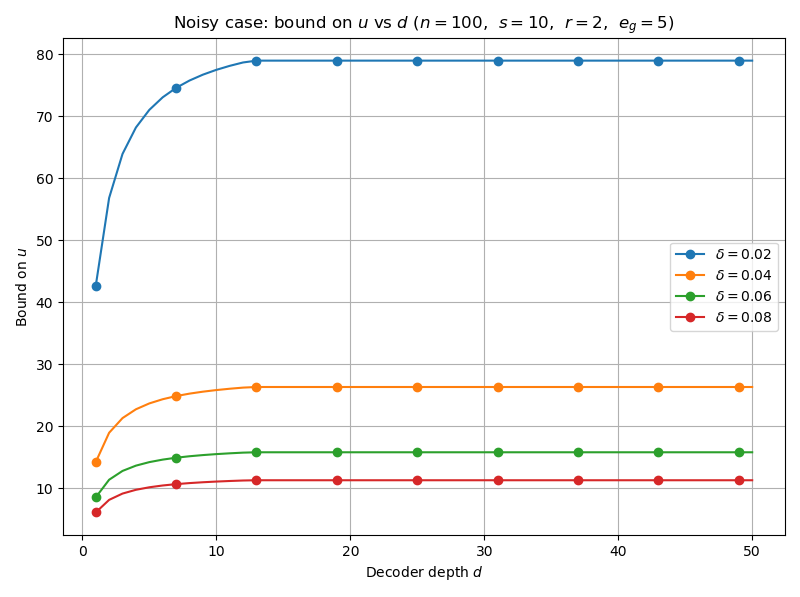} 
	\qquad
	\includegraphics[width=0.4\textwidth]{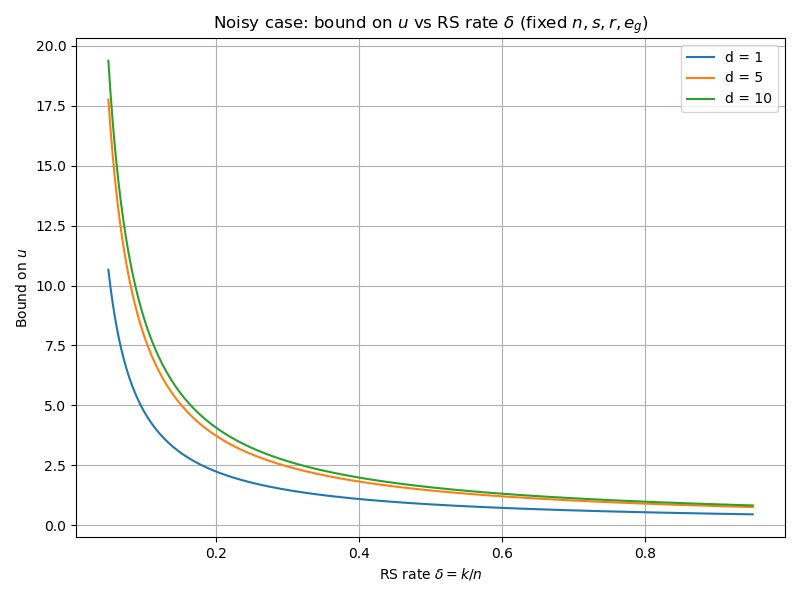}
	\caption{(\textbf{Top}) Bound on $u$ vs $d$ for different values of $\delta = K/N$ (noisy case). (\textbf{Bottom}) Bound on $u$ vs $\delta = K/N$ for different values of $d$ (noisy case).}
	\label{figure:together}
\end{figure}

\ifFULL
In Fig.~\ref{figure:together} (bottom) we provide an illustrative plot of the Reed-Solomon rate~$\delta = K/N$ against the bound on~$u$ in the general noisy case for~$d = 1, 5, 10$, where~$N=100, s=10, r=2, e_g=5$ are fixed for each curve.
One can see that in general, as~$K/N$ decreases, and as the value of~$d$ increases, the capabilities of histogram recovery increases. 
For instance, when~$d = 1$, we can allow only~$\approx 2$ distinct objects in the superposition before recovery is impossible when~$\delta \approx 0.2$, whereas~$\approx 10$ unique superimposed objects can be recovered when~$\delta \approx 0.05$.
When~$\delta \approx 0.05$, one is able to recover~$\approx 10$ distinct objects when~$d = 1$, while the number of recoverable distinct objects rises to about~$18$ when~$d=10$.

In Fig.~\ref{figure:together} (top), we plot the tunable parameter~$d$ against the bound on~$u$ in the noisy case for more fine-grained~$\delta = 0.02, 0.04, 0.06, 0.08$, again fixing~$N=100, s=10, r=2, e_g=5$ for each curve.
We again see the effect that~$\delta$ has on recovery capabilities.
In particular, when~$\delta = 0.08$, we can allow~$\approx 10$ distinct objects in the superposition for successful recovery, whereas~$\approx 80$ distinct objects can be recovered when~$\delta \approx 0.02$.
We may also observe that the effect of~$d$ tends to subside as~$d$ goes above~$10$.
\fi
\ifFULL
\section{Discussion}
In this paper it was shown how coding theoretic techniques give rise to a vector symbolic architecture which subsumes the structural benefits of linear codes~\cite{raviv2024linear}---including efficient encoding, storage, quasi-orthogonality, and binding recovery---in addition to providing an efficient superposition-recovery algorithm by solving the histogram recovery problem. 
\fi
\ifFULL
In what follows we discuss several aspects of our VSA.
\subsection{Noise resilience}
\fi
To summarize, our algorithm guarantees successful recovery of a noisy superposition~$\sum_{i=1}^t\boldc_i+\bolde$ for certain noise vectors~$\bolde\in\bC^{p^{2m}}$.
To state our guaranteed noise resilience, write~$\bolde=(\bolde_1,\ldots,\bolde_N)$, with~$\bolde_i\in\bC^N$ for every~$i$, and for a nonnegative integer~$r$ let~$E_r\triangleq \cup_{\boldv\in \bZ^N\vert \onenorm{\boldv}\le r}\cB_2^o(H\boldv,\sqrt{N}/2)$, where~$\cB_2^o(\cdot,\cdot)$ denotes an open Euclidean ball with a certain center and radius.
The $i$'th Hadamard recovery algorithm in Section~\ref{section:HadamardRecovery} is guaranteed to succeed whenever~$\bolde_i\in E_0$, i.e., $\twonorm{\bolde_i}<\sqrt{N}/2$ (Corollary~\ref{corollary:L2boundHadamard}), while the histogram-recovery algorithm in Section~\ref{section:HistoRecovery} (specifically~\ref{section:notdistinct}) guarantees success even in cases where some of the individual Hadamard recovery algorithm fail.
Hence, we have the following.
\begin{proposition}
	For nonnegative integers~$s,r$ such that~$(r+1)s<N$, our algorithm successfully recovers~$\boldc_1,\ldots,\boldc_t$ from~$\sum_{i=1}^{t}\boldc_i+\bolde$ whenever: (a) the parameter~$t$ obeys Theorem~\ref{thm:distinctbound}; and (b) every~$\bolde_i$ satisfies~$\twonorm{\bolde_i}<\sqrt{N}/2$ except for at most~$s$ many indices~$i$ in which~$\bolde_i\in E_r$.
\end{proposition}
\ifFULL
\begin{proof}
	According to Theorem~\ref{thm:distinctbound} and Section~\ref{section:HadamardRecovery}, it suffices to show that for each one of the~$s$ exceptional~$i$'s with~$\bolde_i\in E_r$, the histogram which results from the Hadamard decoding (Section~\ref{section:HadamardRecovery}) deviates from the correct one by at most~$r$ in~$\ell_1$-distance.
	Then, it would readily follow that all the histograms given as input to the histogram-recovery algorithm are correct, except for at most~$s$ of them which deviate by at most~$r$ in~$\ell_1$-distance, and hence Theorem~\ref{thm:distinctbound} holds.
	
	To this end, let~$\boldh_i$ be the correct histogram for a certain entry~$i$ in which~$\bolde_i\in E_r$, i.e., the input to the Hadamard recovery algorithm~$\bolds_i^\intercal=H\boldh_i^\intercal+\bolde_i^\intercal$.
	By the definition of~$E_r$, there exists~$\boldv\in\bZ^N$ with~$\onenorm{\boldv}\le r$ such that~$\bolde_i\in\cB_2^o(H\boldv,\sqrt{N}/2)$, i.e., there exists~$\bolde_i'\in\cB_2^o(0,\sqrt{N}/2)$ such that~$\bolde_i^\intercal=H\boldv^\intercal+\bolde_i'^\intercal$.
	Therefore, we have~$\bolds_i^\intercal=H\boldh_i^\intercal+H\boldv^\intercal+\bolde_i'^\intercal=H(\boldh_i+\boldv)^\intercal+\bolde_i'^\intercal$. 
	Hence, according to the discussion in Section~\ref{section:HadamardRecovery}, it follows that the rounding algorithm over~$\bolds_i$ outputs the histogram~$\boldh'_i=\boldh_i+\boldv$, and indeed~$\onenorm{\boldh_i-\boldh_i'}=\onenorm{\boldv}\le r$.
\end{proof}

\subsection{Infinite superposition}
Theorem~\ref{thm:distinctbound} readily implies the curious phenomenon that infinitely many objects can be superimposed and recovered successfully, as long as the number of \textit{distinct} objects among them is bounded.
However, superimposing an excessive number of copies of unique objects might interfere with neural retrieval vis-\`{a}-vis~\cite[Thm.~1]{raviv2024linear}, as the latter depends on the \textit{overall} number of objects regardless of uniqueness.
Nevertheless, this phenomenon can be utilized in cases where neural retrieval is not strictly required.
Furthermore, the authors point out that the number of times each object is added onto the superposition may serve as an additional piece of information which does not necessarily reflect the repetition of the object itself.
For instance, VSAs are often used for representing ordered information such as vectors~\cite[Sec.~3.2.2]{raviv2024linear}, in which case the multiplicity of an element in a superposition can serve as its index. 
That is, if the order among objects represented by~$\boldc_1,\ldots,\boldc_t$ is of essence, one can store the superposition~$\sum_{i=1}^t i\cdot\boldc_i$; the number of times each object is stored then serves as its index.

\subsection{Tradeoffs relative to alternative solutions}
Our results drastically improve the use of linear codes in VSAs by presenting explicit and noise-resilient recovery algorithms (the corresponding problems in~\cite{raviv2024linear} are NP-hard).
Yet, this comes at a cost: the random linear codes used in~\cite{raviv2024linear} are of higher rates for a given incoherence level.
Explicit codes which attain similar incoherence to random linear ones were presented in~\cite{ta2017explicit}, but their construction is highly involved, making their deployment in VSAs challenging.
Additional explicit constructions of incoherent linear codes do exist~\cite{wikipedia_epsbalanced}, and finding efficient superposition-recovery algorithm for those is an interesting venue for future work.

Additionally, as mentioned earlier, one may simply use the Hadamard code itself (or its variants~\cite{liu2025linearithmic}) for representing objects---without the additional concatenation operation given in Section~\ref{section:construction}.
This would result in a code of length~$p^{2m}$ and size~$p^{2m}$ (i.e., dimension~$2m$ as a linear code over~$\bF_p$), in which binding-recovery follows from~\cite{raviv2024linear} due to it being linear, and superposition-recovery of infinitely many items follows similar to Section~\ref{section:HadamardRecovery}.
Choosing between this Hadamard code and our~$\cC$~\eqref{equation:cDef} presents a tradeoff: on the one hand~$\cC$ is larger by an exponent of~$K/2$, and on the other hand, a larger~$K$ poses a limitation on the number of superimposed elements that can be efficiently recovered (Theorem~\ref{thm:distinctbound}).
Hence, when choosing between the two in any target application one should weigh the recovery requirements against the anticipated number of objects.
\fi

\printbibliography

\end{document}